\documentclass[aps,prd,reprint,a4paper,floatfix,groupedaddress,amsmath,amssymb,amsfonts,longbibliography]{revtex4-1}
\pdfoutput=1
\usepackage{mathptmx}
\usepackage{textcomp}
\usepackage{helvet}
\usepackage[utf8]{inputenc} 
\usepackage{graphicx}
\usepackage{upgreek} 
\usepackage{xspace}
\usepackage{textcomp}
\usepackage[pdftex]{hyperref}
\usepackage[rightcaption]{sidecap}
\usepackage{color,soul} 
\usepackage{subfigure}
\usepackage[
,textwidth=17.5cm
,textheight=23.5cm
,verbose
,dvips
]{geometry}

\usepackage[colorinlistoftodos,prependcaption,textsize=footnotesize,textwidth=1.5cm]{todonotes}

\begin{document}
\title {The effect of the three-dimensional strain variation on the emission properties of light-emitting diodes based on (In,Ga)N/GaN nanowires}

\author{M. Musolino} 
\affiliation{Paul-Drude-Institut f\"{u}r Festk\"{o}rperelektronik, Hausvogteiplatz 5--7, D-10117 Berlin, Germany}
\author{A. Tahraoui}
\affiliation{Paul-Drude-Institut f\"{u}r Festk\"{o}rperelektronik, Hausvogteiplatz 5--7, D-10117 Berlin, Germany}
\author{L. Geelhaar}
\email[Author to whom correspondence should be addressed. Electronic mail: ]{geelhaar@pdi-berlin.de}
\affiliation{Paul-Drude-Institut f\"{u}r Festk\"{o}rperelektronik, Hausvogteiplatz 5--7, D-10117 Berlin, Germany}
\author{F. Sacconi}
\affiliation{TiberLAB s.r.l., Via del Politecnico 1, 00133 Rome, Italy}
\author{F. Panetta}
\affiliation{TiberLAB s.r.l., Via del Politecnico 1, 00133 Rome, Italy}
\author{C. De Santi}
\affiliation{Department of Information Engineering, University of Padova, Via Gradenigo 6/B, 35131 Padova, Italy}
\author{M. Meneghini}
\affiliation{Department of Information Engineering, University of Padova, Via Gradenigo 6/B, 35131 Padova, Italy}
\author{E. Zanoni}
\affiliation{Department of Information Engineering, University of Padova, Via Gradenigo 6/B, 35131 Padova, Italy}



\begin{abstract} 

In the experimental electroluminescence (EL) spectra of light-emitting diodes (LEDs) based on N-polar (In,Ga)N/GaN nanowires (NWs), we observed a double peak structure. The relative intensity of the two peaks evolves in a peculiar way with injected current. Spatially and spectrally resolved EL maps confirmed the presence of two main transitions in the spectra, and suggested that they are emitted by the majority of single nano-LEDs. In order to elucidate the physical origin of this effect, we performed theoretical calculations of the strain, electric field, and charge density distributions both for planar LEDs and NW-LEDs. On this basis, we simulated also the EL spectra of these devices, which exhibit a double peak structure for N-polar heterostructures, both in the NW and the planar case. In contrast, this feature is not observed when Ga-polar planar LEDs are simulated. We found that the physical origin of the double peak structure is a stronger quantum-confined Stark effect occurring in the first and last quantum well of the N-polar heterostructures. The peculiar evolution of the relative peak intensities with injected current, seen only in the case of the NW-LED, is attributed to the three-dimensional strain variation resulting from elastic relaxation at the free sidewalls of the NWs. Therefore, this study provides important insights on the working principle of N-polar LEDs based on both planar and NW heterostructures. 
  
\end{abstract}

\maketitle

\section{Introduction} 

In the last decade, III-N nanowires (NWs) have been often employed as basis for the fabrication of light-emitting diodes (LEDs)~\cite{Kikuchi2004,Kim2004a,Lin2010,Guo2010,Bavencove2010,Armitage2010,Nguyen2011,Limbach2012a,Ra2013,Musolino2014}. One of the main advantages of this approach is the elastic relaxation of the strain induced by lattice mismatch at the free sidewalls due to NW geometry~\cite{Glas2006,Wolz2013}, thus enabling along the NW axis  the growth of high quality (In,Ga)N/GaN heterostructures with In content higher than in conventional planar LEDs. Because of strain relaxation effects, the strain distribution inside the active region of LEDs based on NWs significantly differs from the one present in planar devices~\cite{Wolz2013}. This effect can substantially influence the optoelectronic properties of the NW-LEDs~\cite{Marquardt2013a}. The strain distribution is particularly relevant for III-nitride semiconductors, where the strain not only shifts the conduction and valence band edges, but also produces piezoelectric fields leading to a redistribution of charge carriers~\cite{Bernardini1997,Marquardt2013a}. Despite the significant difference in strain distribution between LEDs based on NWs and on planar layers, the consequences of this intrinsic aspect for the emission properties have not been studied in comparison.

In this work, we carefully investigate the optoelectronic properties of green N-polar III-N NW-LEDs experimentally as well as theoretically, and we compare the main features of LEDs based on planar layers and NWs. To this end, we measured the electroluminescence (EL) of a NW-LED fabricated from a NW ensemble as a function of current. In these EL spectra, we observe a double peak structure, and the relative intensity of the two peaks exhibits a peculiar evolution with injected current. Spatially and spectrally resolved EL maps confirm the presence of two main lines in the spectra, and suggest that they are emitted by the majority of NWs in the ensemble LED. 

In order to understand the physical origin of this effect, we performed theoretical calculations of strain, electric field, and charge carrier distributions inside the active region of both planar LEDs and NW-LEDs, each one containing four quantum wells (QWs). On this basis, we simulated the EL spectra of the two types of devices. We found that the physical origin of the double peak structure is a stronger quantum-confined Stark effect (QCSE) in the first and last QW of the heterostructure with four QWs. This phenomenon results from the nitrogen crystal polarity and is observed as well in the simulated EL spectra of planar N-polar LEDs. The simulations show also that the peculiar evolution of the relative EL peak intensities with injected current, observed only in case of the NW-LED, is caused by the strain relaxation characteristic for NWs. Therefore, we directly identify consequences of the NW morphology on the emission properties of LEDs. Furthermore, we provide important insights on the recombination mechanisms of N-polar LEDs based on both planar and NW heterostructures.

\section{Methods} \label{sec:methods}

\begin{figure} [b]
\includegraphics*[height=7.1cm]{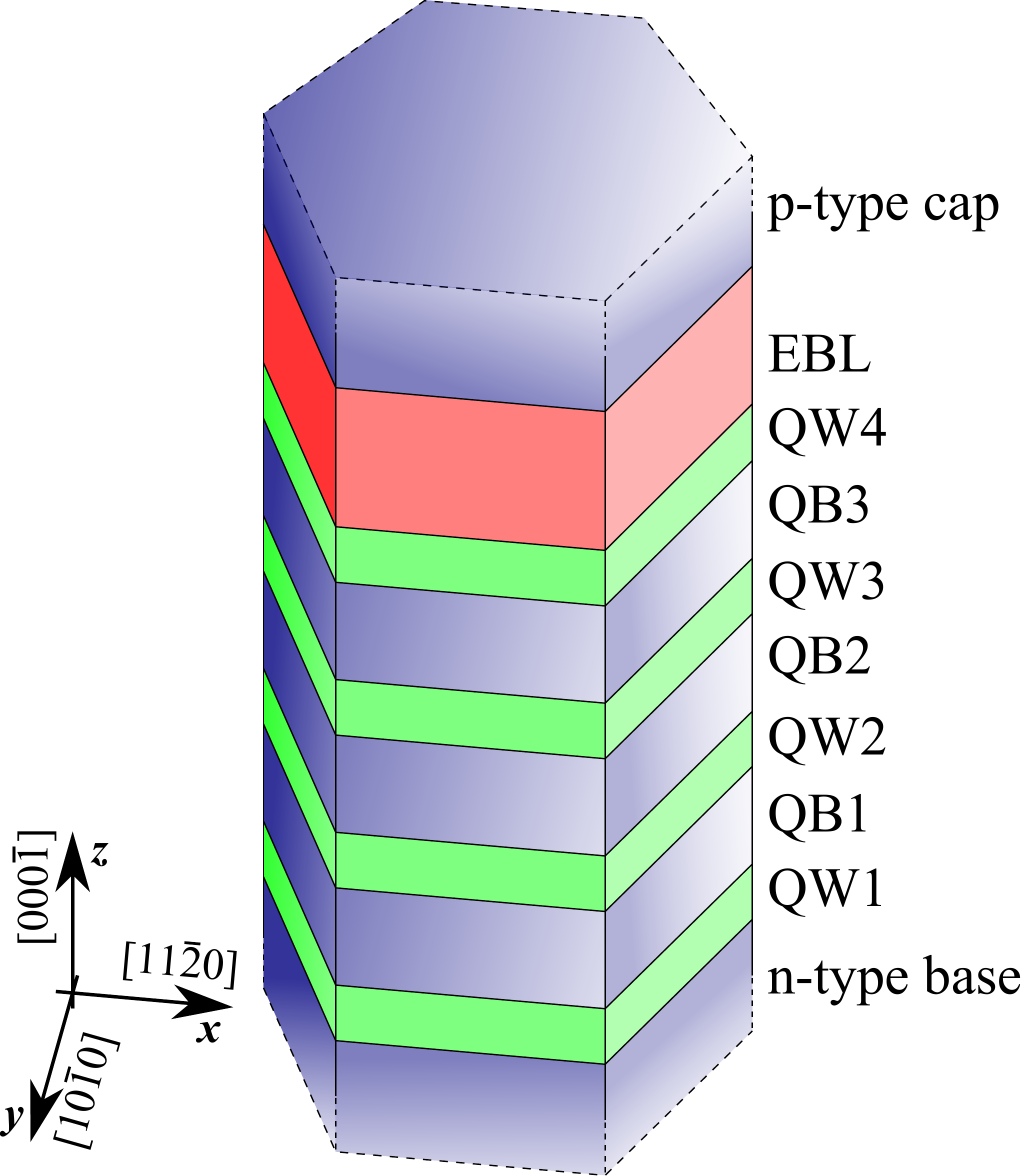} 
\caption[]{Illustrative sketch of the active region of a NW-LED. Note that the different dimensions are not to scale. The crystallographic directions are shown on the bottom-left corner.}
  \label{fig:sketch}
\end{figure}

The NW-LED structure employed in this work was grown by molecular beam epitaxy on an AlN-buffered n-doped Si(111) substrate with the help of self-assembly processes. Figure~\ref{fig:sketch} depicts an illustrative sketch of the active region employed for the NW-LEDs. It consists of four (In,Ga)N QWs with an average In content of $(25\pm 10)\%$ and a nominal thickness of 3\,nm, separated by three GaN quantum barriers (QBs). The In content in the QWs was estimated through the analysis of the In desorption from the sample during the growth, measured \emph{in situ} by means of line-of-sight quadrupole mass spectrometry~\cite{Wolz2012b,MusolinoPhD}. The last quantum well (labelled QW4) is immediately followed by a Mg-doped (Al,Ga)N electron blocking layer (EBL) with a nominal Al content of about 25$\%$. Note that such self-assembled GaN NWs are N-polar~\cite{Hestroffer2011,Fernandez-Garrido2012}. The intrinsic (i-type) active region is embedded between two doped GaN segments, namely the n-type (Si-doped) base and the p-type (Mg-doped) cap, designed such that an n-i-p diode profile is created. For device fabrication, the space between NWs was filled with an insulator, and a transparent top contact as well as a back contact were deposited. The NW-LED device investigated in this study has an area of 0.19\,mm$^2$. 
More details about the employed LED structure and growth parameters as well as the fabrication process can be found in our previous publications~\cite{Musolino2014,Musolino2015a,Musolino_NT_2016}. 

The EL spectra of the LED device were acquired at room temperature ($T=303$\,K) in a dark ambient; the light was collected through an optical microscope and analyzed by an array spectrometer with a CCD camera. In addition, spectrally resolved EL top-view maps of the same NW-LED were acquired by means of an electrically tunable liquid crystal filter with a bandwidth of about 7\,nm. The spatial resolution of this technique is given by the magnification of the objective and by the density of pixels of the CCD camera employed to record the EL maps. An objective with 100$\times$ magnification was chosen so that each EL spot, whose size is limited by the diffraction of the lens to about 0.2\,$\mu$m,  is recorded simultaneously by several pixels of the CCD camera. 

One- (1D) and three-dimensional (3D) theoretical simulations of spatial distributions of strain, electric field, and charge carrier density as well as EL spectra of LED structures similar to the one shown in Fig.~\ref{fig:sketch} were performed by means of the software TiberCAD~\cite{AufderMaur_2011}. This simulation tool employs a parametric multiscale coupling approach. Namely, material parameters extracted from first-principle atomistic calculations are used as input for high level continuous media simulations. In particular, we focus on the composition dependence of band gap as well as conduction and valence band edge bowing parameters for the (In,Ga)N alloy. Commonly, the band gap bowing parameter of (In,Ga)N is assumed to be composition independent~\cite{Vurgaftman2003}. However, more recent results by Caro \emph{et al.}~\cite{Caro2013}, based on empirical tight-binding random supercell calculations, showed a strong composition dependence of this parameter. The results of Caro \emph{et al.} were implemented in the material database of TiberCAD~\cite{Sacconi2015} and used for the simulations presented in this work. 

In order to simulate the optoelectronic properties of the LEDs, elasticity models were applied to obtain strain maps and piezoelectric polarization. The effects of strain on band edges were included through deformation potentials~\cite{Yan2009,Li2003}. Then, the LED transport properties and I-V characteristics were calculated by means of a drift-diffusion model, which includes the total polarization fields. Finally, eigenstates and eigenfunctions were calculated in the framework of a 6$\times$6 k$\cdot$p model in an envelope function approximation for several injected current densities~\cite{AufderMaur2015}. On this basis, optical spectra for spontaneous emission were calculated through the optical matrix elements between conduction and valence states. The emitted power spectral density was then determined by evaluating the optical transition probability weighted with the occupation functions for each couple of particles and energy state. More details about the simulation method can be found elsewhere~\cite{AufderMaur_2011,AufderMaur2015}.

For the simulation of single nano-LEDs, a diameter of the column equal to 80$\,$nm was chosen; while QWs, QBs, and EBL thicknesses were set to 3, 8, and 15\,nm, respectively. The In and Al contents inside QWs and EBL were varied in the ranges $[10\--32]\%$ and $[15\--25]\%$, respectively. The best agreement with the measured EL spectra was obtained using an In content of 32$\%$ for each QW and an Al content of 25$\%$ in the EBL, values which are in good agreement with the experimental estimations. Concentrations of donors and acceptors equal to $3\times 10^{19}$\,cm$^{-3}$ and $7\times 10^{18}$\,cm$^{-3}$, respectively, were chosen.

Note that the real structure of the NWs may be more complex than that depicted in Fig.~\ref{fig:sketch}. For example, it was observed in similar NWs that the (In,Ga)N insertions may be surrounded by GaN laterally, both their top and bottom interface may not be flat. In addition, the In content may vary from insertion to insertion as well as within each insertion~\cite{Tourbot2012,Woo2015}. Furthermore, any NW ensemble grown on the basis of self-assembly processes is prone to ensemble fluctuations. At the same time, a comprehensive description of all these details would be unrealistic, in terms of both experimental characterization and theoretical simulation. Thus, we focus on the basic structure in the simulations and verify their validity by comparison with the experiments. Similarly, many defects characteristical of planar (In,Ga)N/GaN heterostructures, such as threading and misfit dislocations as well as In clusters, were not considered in the 1D simulations of the LED.

\section{Experimental observations}

\begin{figure} [b]
\includegraphics*[width=8cm]{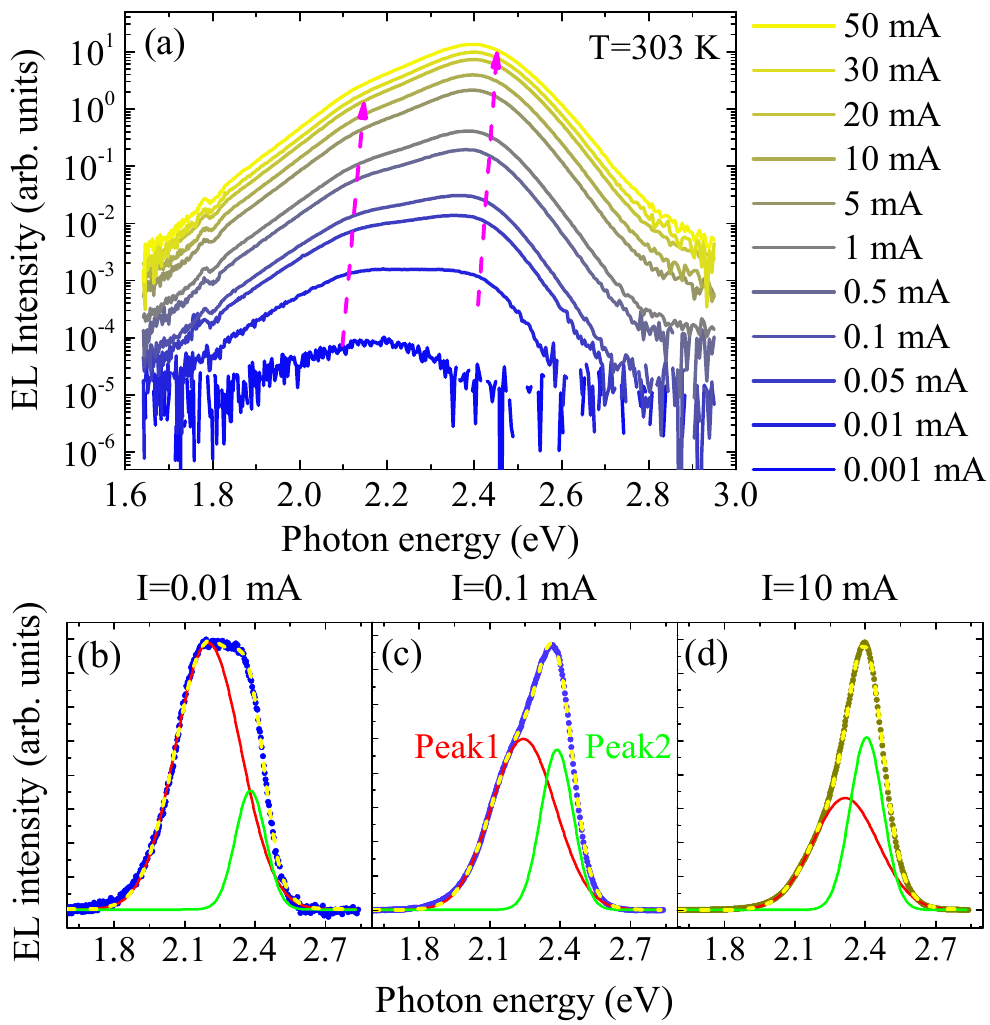} 
\caption[]{(a) EL spectra of the NW ensemble for total injected current from 0.001 to 50\,mA, plotted on a semi-logarithmic scale. The magenta arrows are guides to the eye, which sketch the evolution of the two main transitions. (b)$\--$(d) Exemplary EL spectra (data points) and fits obtained by means of two Gaussians labelled “Peak1” and “Peak2”, see red and green solid curves, respectively. The yellow dashed lines depict the cumulative fits, namely the algebraic sum of the two Gaussians. The data are plotted on a linear scale and refer to injected currents of (b) 0.001\,mA, (c) 0.1\,mA, and (d) 10\,mA.}
  \label{fig:EL-I}
\end{figure} 

Figure~\ref{fig:EL-I}(a) shows the evolution of the EL spectra of the NW ensemble for injected currents ranging from 0.001\,mA up to 50\,mA. The EL spectra are characterized by a rather broad emission band consisting of two main transition lines, which exhibit a blue shift with increasing injected current, as marked by the magenta dashed arrows. Interestingly, similar asymmetric peaks with low energy tails have also been observed in other green NW-LEDs reported in literature with structures comparable to the one employed in this work~\cite{Bavencove_nt_2011,Limbach2012a,Kishino2014a,Musolino2014}. However, only the integrated EL or the position of the most intense transition has been considered in all these cases. In this work, we aim at studying the evolution of the different lines constituting the emission band in more detail. To this end, the EL spectra were fitted by means of two Gaussians. The resulting fits for three different currents are plotted in Figs.~\ref{fig:EL-I}(b)$\--$(d). Note that the cumulative fits, depicted by the yellow dashed lines, perfectly overlap the data points. The relative intensities of the two transitions, labelled “Peak1” and “Peak2”, evolve in a peculiar way. At very low current ($I=0.01$\,mA) the low energy line (Peak1) is much more intense than the high energy line (Peak2), but the latter increases faster and, for $I=0.1$\,mA, it has already an intensity comparable with the one of Peak1. At higher currents the high-energy line overtakes the low-energy one and dominates the spectrum.

\begin{figure}[b]
\includegraphics*[width=8cm]{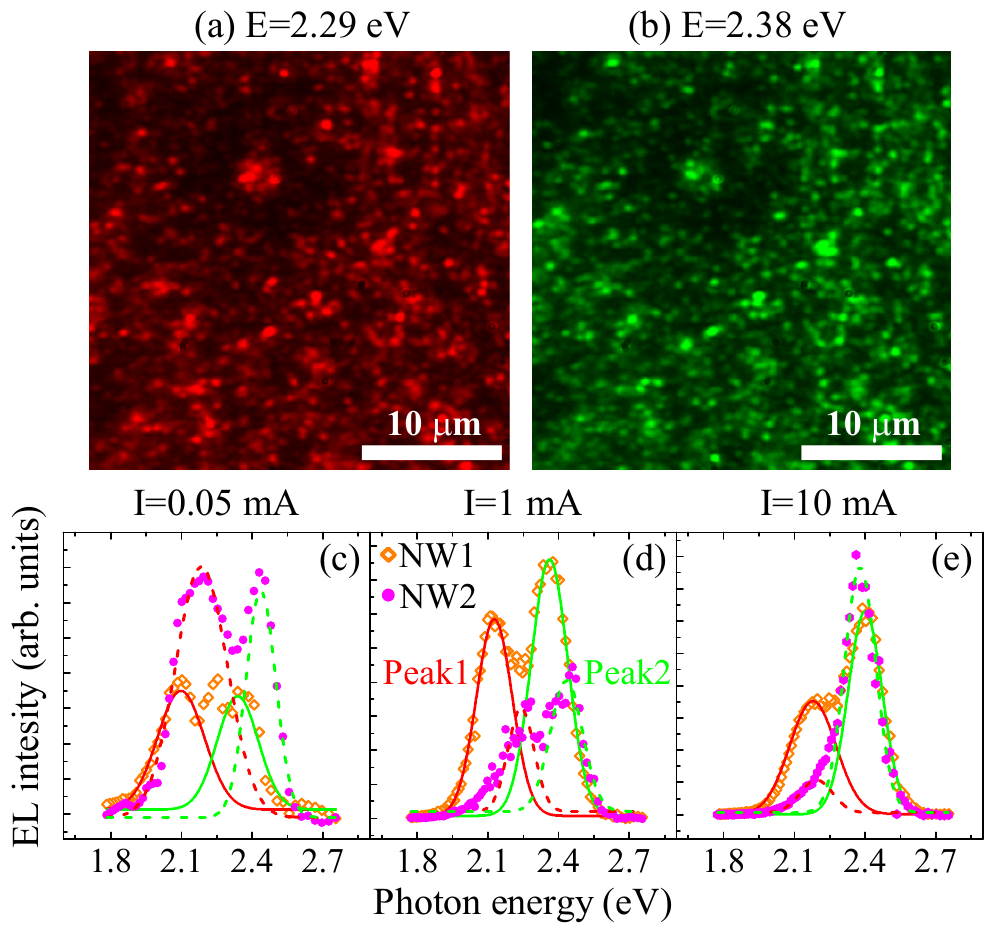} 
\caption{Monochromatic false-color top-view EL maps of the NW-LED for emission energies of (a) 2.29 and (b) 2.38\,eV acquired under a total injected current of 20\,mA. The emission energies correspond to the two main peaks observed in the EL spectra of Fig.~\ref{fig:EL-I}. (c)$\--$(e) Spectra extracted from two different single emission spots on the EL maps, attributed to individual NWs (NW1 and NW2), for injected currents of 0.05, 1, and 10\,mA. The solid and dashed lines represent the two Gaussians used to fit the data points of NW1 and NW2, respectively.}
\label{fig:EL-I-map}
\end{figure}

It is important to understand if the two transitions occur in the majority of the individual NW-LEDs or if they are due to a cumulative effect of the NW ensemble produced by a superposition of the emission spectra of various NWs emitting at different energies. To answer this question, spectrally resolved EL top-view maps of the same NW-LED were acquired. The resulting monochromatic EL maps for the emission energies corresponding to the two main transitions of 2.29 and 2.38\,eV are shown in Figs.~\ref{fig:EL-I-map}(a) and (b), respectively. A comparison of the two EL maps shows that almost each emission spot, which likely corresponds to the emission of a single nano-LED or a cluster of few NWs, is present at both emission energies. This fact strongly suggests that the two transitions observed in the EL spectra of the NW ensemble are actually related to common properties of the individual NWs rather than to ensemble inhomogeneities. 

The EL maps were acquired for several injected currents in the spectral range from 1.8 to 2.8\,eV. From the EL maps at different energies, it is possible to extract a spectrum for each pixel of the map and each injection current. Thus, EL spectra related to different emission spots can be analyzed in detail. Several points were checked, but for sake of clarity only the typical spectra related to two of them, hereafter named NW1 and NW2, are plotted in Figs.~\ref{fig:EL-I-map}(c)$\--$(e). Although the EL intensity exhibits significant fluctuations from spot to spot, the majority of the spectra are characterized by two main transitions while the energy positions of these lines differ slightly between different NWs. This finding confirms the presence of fluctuations in the emission properties of single nano-LEDs already observed by different groups~\cite{Bavencove_nt_2011,Lahnemann2014}. Again, we fitted the spectra by means of two Gaussians (Peak1 and Peak2). Figures~\ref{fig:EL-I-map}(c)$\--$(e) show the representative spectra acquired at three different currents injected into the NW ensemble (0.05, 1, and 10\,mA). The same peculiar evolution of the EL intensity with increasing current observed for the ensemble measurements is visible also in the spectra from single emission spots. For small currents [see Fig.~\ref{fig:EL-I-map}(c)], the low energy line has higher or similar intensity than the high energy one. With increasing current, Peak2 rises faster than Peak1 and eventually dominates the spectra [see Figs.~\ref{fig:EL-I-map}(d) and (e)]. A similar behavior was observed in almost all analyzed EL spots, which confirms that the emission of the majority of the individual NWs is characterized by two main lines.

\section{Simulations of the optoelectronic properties}

The experimental observation of two main transitions in the EL spectra of NW-LEDs, characterized by a high energy line whose intensity dominates at higher injected currents, conflicts with the shape of the EL spectra emitted by conventional LEDs based on planar (In,Ga)N/GaN heterostructures. In fact, their EL spectra exhibit only one asymmetric peak without significant variation of its shape with increasing injected current~\cite{Schubert_Book,MorkocBook3}. It is tempting to attribute this difference to the additional strain relaxation present in NWs. At the same time, commercial GaN-based LEDs differ in many aspects from NW-LEDs: the planar morphology in place of the 3D one, the Ga-polarity of the crystal structures used in conventional LEDs instead of the N-polarity typical of self-induced NW-LEDs~\cite{Hestroffer2011,Fernandez-Garrido2012}, the rather different growth methods and processing schemes, etc. Thus, the origin of the double peak structure observed in our EL spectra is not obvious. 

In order to understand the difference between planar and NW-LEDs, we simulated the optoelectronic properties of Ga- as well as N-polar planar LEDs and NW-LEDs using 1D and 3D models, respectively. Note that for the two types of simulations the same structural and material parameters (namely, QW thickness, In and Al concentrations, doping, etc.) were used.

\subsection{Spatial distribution of strain and electric field}

\begin{figure}[t]
\includegraphics*[width=8cm]{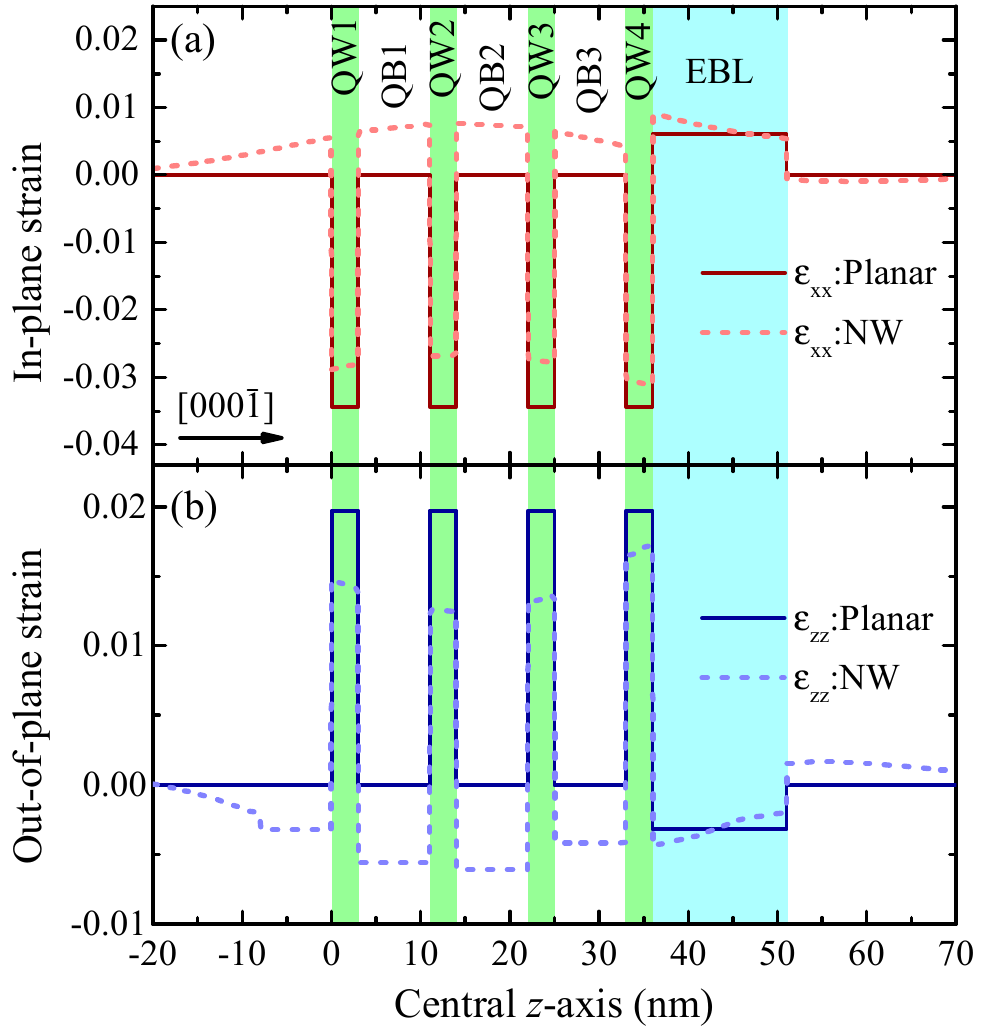} 
\caption{Simulation of (a) in-plane and (b) out-of-plane strain distribution along the central $[000\bar{1}]$ $z$-axis of the employed LED structure. Solid and dashed lines represent the planar and NW case, respectively. The origin of the central $z$-axis is placed at the lower interface of QW1.}
\label{fig:strain}
\end{figure}

Figures~\ref{fig:strain} (a) and (b) show the in-plane and out-of-plane strain distribution along the central $[000\bar{1}]$ $z$-axis of the LED structure, which is depicted in Fig.~\ref{fig:sketch}, for both the planar and NW-based N-polar devices. Note that both in-plane and out-of-plane strain components, represented by $\epsilon_{xx}$ and $\epsilon_{zz}$ respectively, are constant inside the different parts of the planar LED, whereas they vary in the NW-LED. Inside the four QWs, a compressive (\textit{i.\,e.}, negative) in-plane strain is present, which results in tensile (\textit{i.\,e.}, positive) strain in the out-of-plane direction. Note that, the absolute values of the strain inside all the QWs is lower in the NW case than in the planar one. Furthermore, the QBs of the planar LED are not strained, whereas a significant strain is induced in the barriers of the NW-LED. The more complex strain distribution inside the various parts of the active region, and the lower absolute values of the strain in the NW-LED are direct consequences of the strain relaxation occurring at the free sidewalls of the NWs. This effect is not possible in planar heterostructures. We highlight that the strain results obtained in this work are in good agreement with the calculations performed by W\"olz \textit{et al.} for similar structures~\cite{Wolz2013}. Note that, in both the planar and the NW case, the presence of an EBL breaks the symmetry of the active region introducing a total strain at the interface with QW4 higher than the one present in the other QWs. 

\begin{figure}[t]
\includegraphics*[width=8cm]{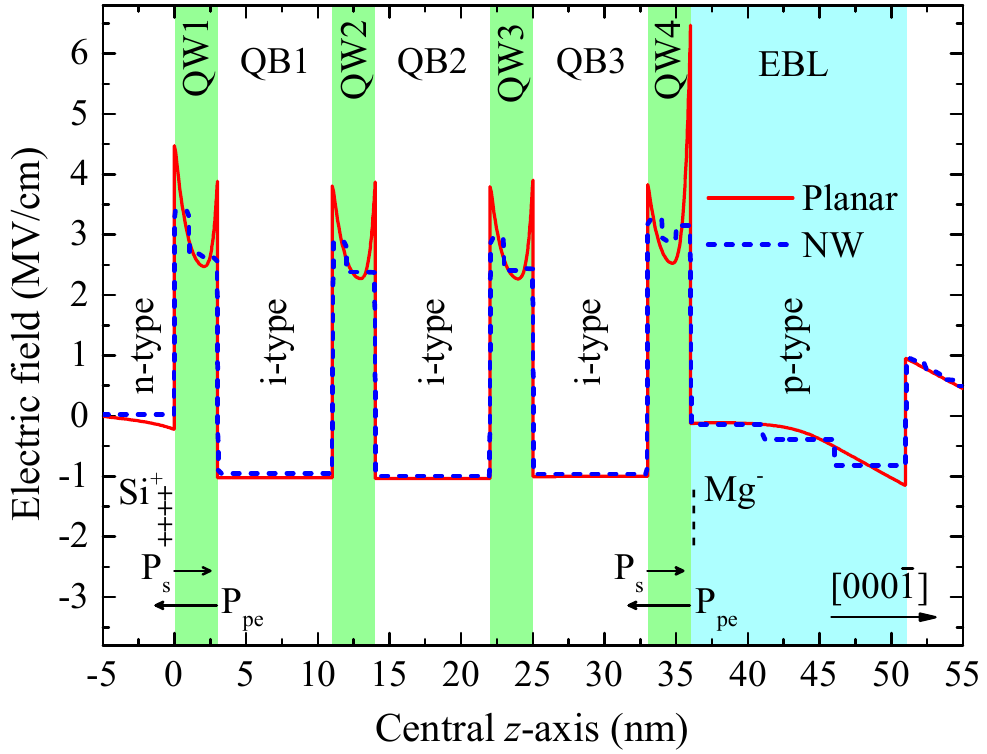} 
\caption{Simulation of the electric field distribution along the central $[000\bar{1}]$ $z$-axis of the employed LED structure calculated for an injected current density of 180\,Acm$^{-2}$. Solid and dashed lines depict the data for the planar and NW case, respectively. The vectors represent direction and magnitude of spontaneous polarization (\textbf{P$_{s}$}) and piezoelectric polarization (\textbf{P$_{pe}$}). The positive or negative charge accumulation due to ionized donors and acceptors, namely Si$^+$  and Mg$^-$ atoms, is indicated by "++++" and "-\,-\,-\,-"", respectively. The origin of the central $z$-axis is placed at the lower interface of QW1.}
\label{fig:field}
\end{figure}

In a piezoelectric material, such as GaN, the strain contributes to the accumulation of net charge at the interfaces between the different sections of the heterostructure, which is known as piezoelectric polarization (\textbf{P$_{pe}$}). In addition, in the wurtzite crystal structure, atoms of opposite electronegativity alternate along the $[0001]$ direction. This fact causes a charge rearrangement leading to a dipole momentum along this axis, named spontaneous polarization (\textbf{P$_{s}$}). In N-polar QWs under compressive strain, the vectors \textbf{P$_{pe}$} and \textbf{P$_{s}$} have opposite direction, as shown by the black arrows in Fig.~\ref{fig:field}. The total polarization, namely the vectorial sum of \textbf{P$_{pe}$} and \textbf{P$_{s}$}, generates an electric field in the same direction of the built-in electric field. 

Under a steady state condition, the positive and negative net charges induced on the opposite sides of each QW by piezoelectric and spontaneous polarization are in equilibrium with the charges in the adjacent regions. These are, for instance, the positive ionized donors or the negative ionized acceptors in the n-type and p-type side, respectively. Moreover, in forward biased LED devices, additional electrons and holes are injected into the QWs from the n-type and p-type side, respectively. The rearrangement of all these charges on the opposite sides of the various sections of the heterostructure determines the electric field distribution inside the active region of the LED.

Figure~\ref{fig:field} shows the simulated electric field distribution along the central $[000\bar{1}]$ $z$-axis of the employed LED structure calculated for an injected current density of 180\,Acm$^{-2}$. Note that magnitude and distribution of the electric field are essentially identical in QW2 and QW3 in both the planar and NW case. In contrast, the electric field distribution inside QW1 and QW4 differ both between them and from the one present inside the two intermediate QWs. In particular, the electric field is stronger in QW1 and QW4 than in QW2 and QW3. The stronger electric field inside QW1 is caused by an asymmetric net charge concentration at the interface between the n-type base and the i-type QW1 due to the different doping levels in the two regions. In fact, the positive charge of the ionized donors (\textit{i.\,e.}, Si$^+$ atoms) located at the end of the base increases the positive charge already present on the bottom side of QW1. A similar effect, caused by the negative charge of the ionized acceptors (\textit{i.\,e.}, Mg$^-$ atoms), is present in QW4, but not in QW2 and QW3, which are enclosed by i-type QBs. 

Furthermore, we highlight that the electric field inside the QWs is lower in the NW case than in the planar counterpart, particularly at the interfaces. This difference is explained by the strain relaxation at the free sidewalls of the NWs that reduces the magnitude of the piezoelectric polarization, and in turn lowers the QCSE with respect to planar LEDs.

\subsection{EL spectra}

\begin{figure*}[ht]
\begin{subfigure}{}
  \centering
  \includegraphics[width=8.5cm]{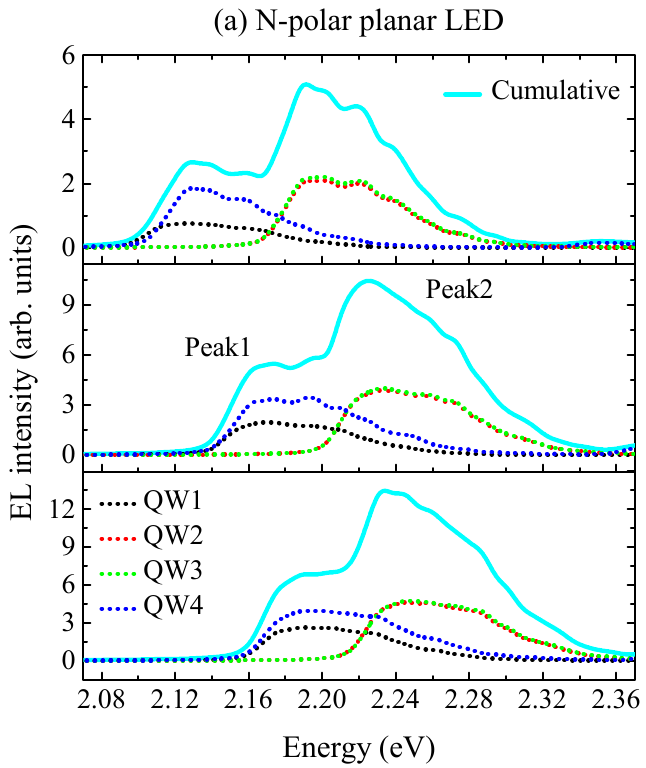}
\end{subfigure}
\begin{subfigure}{}
  \centering
  \includegraphics[width=8.5cm]{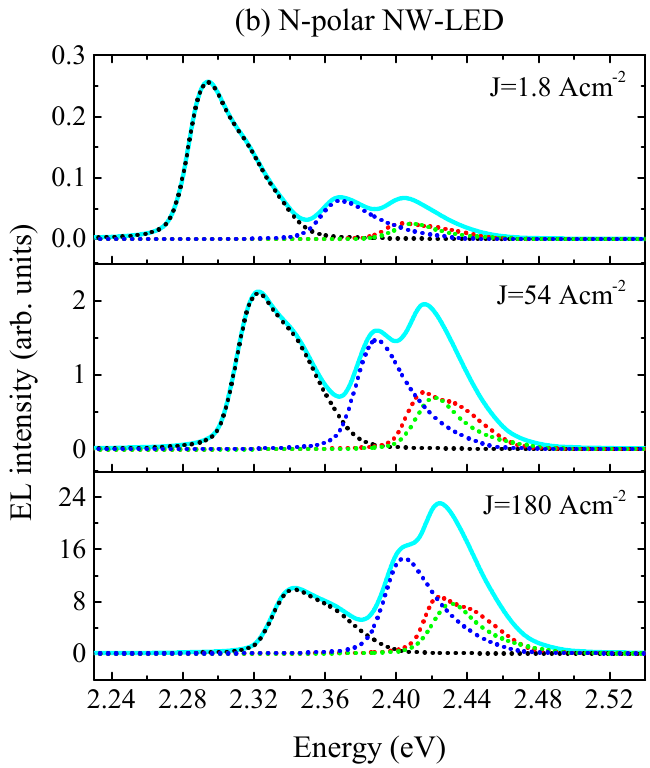}
\end{subfigure}
\caption{Simulation of room temperature EL spectra in case of (a) N-polar planar LEDs and (b) N-polar NW-LEDs calculated  for three different injected current densities (J): 1.8, 54, and 180\,Acm$^{-2}$. The dotted lines represent the emission from each single QW, whereas the solid lines are the cumulative spectra, namely the sum of the different contributions.}
\label{fig:spectra}
\end{figure*}

The distribution of the electric field in the active region defines the magnitude of the QCSE, which causes a shift of the emission energy towards lower values~\cite{Lefebvre_prb_2004}. In order to assess these consequences in detail, we calculated the EL spectra of both planar and NW-LEDs. Localization effects arising from fluctuations in In content~\cite{Lahnemann2014} were not taken into account. Furthermore, the presence of deep-level traps, whose effect can be important in GaN-based LEDs, both in the planar and NW case~\cite{Meneghini_2012,Meneghini2014,Musolino2016a}, was neglected. Possible limitations of this simplified description are discussed in section~\ref{sec:Comp_exp_sim}.

Figures~\ref{fig:spectra}(a) and (b) show the simulated EL spectra for N-polar planar LEDs and NW-LEDs, respectively, calculated for three different injected current densities at room temperature. The dotted lines represent the emission from each single QW, whereas the solid lines are the cumulative spectra. The convolution of the EL emitted by the four QWs is characterized by two main transitions labelled as “Peak1” and “Peak2” both in the planar and NW case. In the planar case, QW2 and QW3 exhibit almost identical emission properties and are responsible for the formation of Peak2, whereas the EL from QW1 and QW4 is red-shifted by about 60\,meV from Peak2, thus creating the low-energy peak. The red-shift of the emission energies of QW1 and QW4 can be explained by the stronger QCSE induced by the higher electric field present inside these two QWs (see Fig.~\ref{fig:field}). Note that Peak1 and Peak2 have similar relative intensities for all the investigated current densities, varying over two orders of magnitude. Thus, the simulations of planar LEDs do not reproduce the peculiar evolution of the relative EL peak intensities with current experimentally observed for NW-LEDs.

The simulated EL spectra of the NW-LED as shown in Fig.~\ref{fig:spectra}(b) exhibit also two main transitions. However, the low energy line originates exclusively from QW1. The emission energy of QW4 is closer to the energies of QW2 and QW3 than to the one of QW1, thus contributing to the broad Peak2. The red-shift of the EL emitted by QW4, which is weaker than in the planar case, can be explained by the weaker electric field in the fourth QW of the NW-LEDs. In fact, Fig.~\ref{fig:field} shows that the strength of the electric field in QW4 is about twice as large for planar LEDs as for NW-LEDs. We attribute this effect to the strain relaxation occurring at the sidewalls of the NWs. Since three QWs contribute to the high-energy peak, the EL intensity of Peak2 increases faster with increasing current than the intensity of Peak1. This behaviour reproduces, at least qualitatively, the evolution of the EL spectra with increasing current observed experimentally. 

\subsection{Spatial distribution of charge carriers}

To fully understand the EL spectra of the NW-LEDs, considerations based only on QCSE and electric field distribution along the central $z$-axis are not sufficient. In particular, the high brightness of QW1 at low injected carrier levels and the rapid increase of the EL intensity emitted by QW4 with current can hardly be explained by the conventional QCSE alone. Therefore, we analyzed the 3D distribution of the electron and hole densities inside each QW of the NW-LED as a function of the injected current. 

\begin{figure*}[t]
\includegraphics*[width=17.5cm]{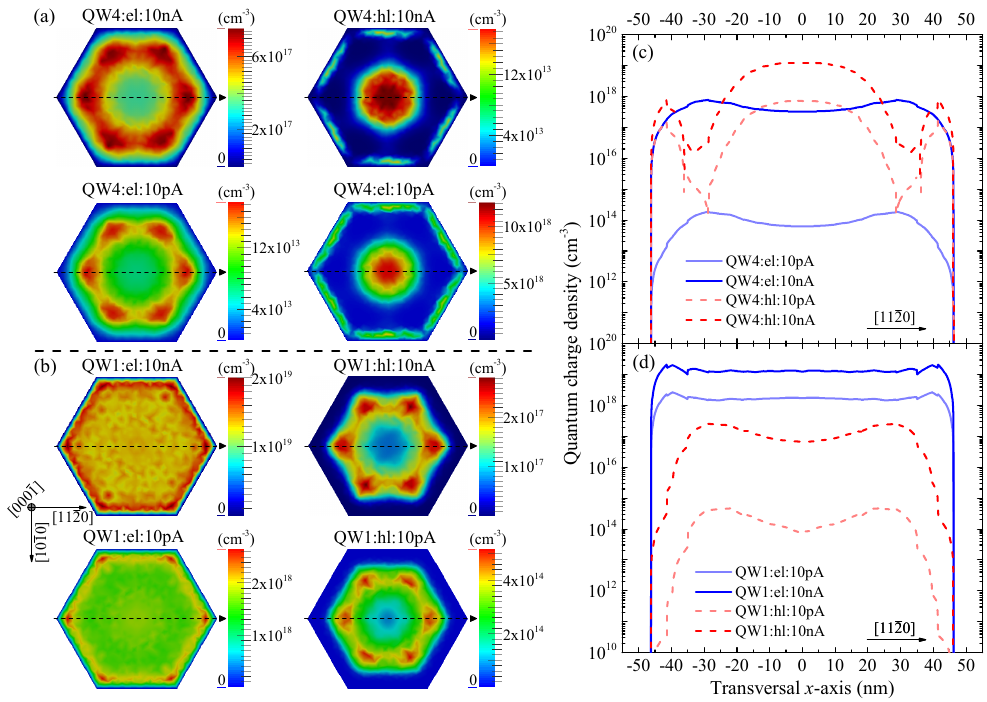} 
\caption{Simulations of the quantum charge density maps on the \textit{x-y} plane of the NW, calculated for (a) QW4 and (b) QW1 at two different injected currents: 10\,pA and 10\,nA. The maps on the left-hand side refer to electrons (el), whereas the ones on the right-hand side to holes (hl). The intensity is color-coded as indicated by the scale bars. Profiles of the quantum charge density, (c) and (d), along the $x$-axis, namely the $[11\bar{2}0]$ direction, extracted from the charge density maps in (a) and (b) along the dashed arrows for QW4 and QW1, respectively. Solid and dashed lines depict the profiles of electron and hole densities, respectively. The origin of the $x$-axis is placed at the centre of the NW.}
\label{fig:charge_maps}
\end{figure*}

Figures~\ref{fig:charge_maps}(a) and (b) depict the 2D top-view maps of electron (el) and hole (hl) quantum charge densities obtained from transversal cross sections of the 3D distributions of charge carriers inside the QWs. The maps were calculated for injected currents of 10\,pA and 10\,nA for each QW (only the data related to QW1 and QW4 are shown). The cross sections were taken on the \textit{x-y} plane perpendicular to the central $z$-axis of the NW [\emph{i.\,e.}, the $(000\bar{1})$ planes] on which the quantum charge densities reach their maximum. These planes have slightly different heights (\emph{i.\,e.}, values of $z$) for electrons and holes. In fact, because of the polarization fields, electrons tend to be confined closer to the bottom part of the QWs, whereas holes are closer to the upper part. Along the $[000\bar{1}]$ direction, the maxima of electron and hole distributions inside each QW are separated  by a distance of about 1\,nm.     

The charge density maps in Figs.~\ref{fig:charge_maps}(a) and (b) show that both electrons and holes inside the QWs of the NW-LED are not uniformly distributed on the \textit{x-y} plane. Furthermore, charge arrangement is significantly different in the two QWs. In particular, the electron density is higher in a ring-shaped region, while holes are localized at the centre and close to the edges of the NW inside QW4. In contrast, electrons are more uniformly distributed and holes tend to accumulate in a ring-shaped region inside QW1. 

Carrier localization phenomena inside (In,Ga)N QWs are often explained taking into account In segregation effects~\cite{Jeong2015}, strain fluctuations induced by defects~\cite{Lahnemann2014}, or also the interplay of surface and polarization potentials~\cite{Marquardt2013a}. In our model, In atoms are supposed to be uniformly distributed. Furthermore, neither defects in the crystal lattice nor potentials induced by surface states are considered. Therefore, none of the previous effects should be responsible for the in-plane localization observed in our simulations. The charge localization as shown in Fig.~\ref{fig:charge_maps} is attributed to the complex strain distribution inside the NW-LEDs. This conclusion is also supported by the clear difference in the charge density maps of the various QWs, which differ mainly in their strain distribution. 

We highlight the fact that the observed in-plane charge localization is a fundamental difference between the planar and NW case. In fact, in the absence of crystal defects or In segregation, an ideal planar (In,Ga)N QW should have a uniform charge distribution on the \textit{x-y} plane, at least on the size-scale considered in our analysis (namely, several tens of nanometres). Note that the in-plane strain is supposed to be homogeneous in a planar (In,Ga)N QW. Therefore, a profile of the charge density along the $z$-axis would be sufficient to describe the emission properties of an ideal planar LED. This is not true in the case of NW-LEDs, even in the ideal case of QWs free of defects and devoid of In segregation effects.    

Figures~\ref{fig:charge_maps}(c) and (d) show the profiles of the quantum charge density along the $[11\bar{2}0]$ direction extracted from the top-view charge density maps for QW4 and QW1. The rapid increase of the EL intensity emitted by QW4 with current as shown in Fig.~\ref{fig:spectra}(b) can be explained by the evolution of the charge density profile as presented in Fig.~\ref{fig:charge_maps}(c). We note that the electron density is rather small at low current, hence the radiative recombination rate $R$ is small as well, thus resulting in a weak EL emission. We recall that $R=Bnp$, where $B$ is the bimolecular coefficient, while $n$ and $p$ represent the electron and hole concentration, respectively. At higher current, the electron density inside QW4 rises significantly, becoming comparable to the hole density in a wide spatial region. The fast rise of the electron density with current and its presence in spatial regions where also the hole concentration is high produces a rapid increase of the radiative recombination rate with a consequent rise of the EL intensity, as observed in the simulated EL spectra.

The 2D charge distribution also explains the high EL intensity emitted by QW1 at low current. Figures~\ref{fig:charge_maps}(b) and (d) show that a rather high electron density is present basically everywhere on the \textit{x-y} plane at low injected current. This leads to a high recombination rate all over the QW, thus resulting in a significant EL emission. In other words, while in QW4 electrons and holes are localized in different spatial regions on the \textit{x-y} plane, in QW1, the electrons are weakly localized and can recombine more easily with holes. We attribute these differences to the non-uniform 3D strain distribution inside the QWs, which induces a complex redistribution of electric fields and charge carriers.

\subsection{Ga-polar \textit{vs.} N-polar LEDs}


The analysis of the simulated EL spectra suggests that the strain relaxation induced by the 3D morphology of the NWs is responsible for the current dependence of the EL spectra, but does not cause the double peak structure. In fact, the latter is observed also in the simulation of N-polar planar LEDs, and is essentially a consequence of the QCSE. 

To further clarify the origin of the double peak structure, which is usually not observed in conventional LEDs based on Ga-polar heterostructures, we compared simulations of the EL spectra from two planar LED structures that differ only in the crystal polarity (see Supplemental Material~\cite{Supplemental_Material}). The calculations have shown that, in contrast to N-polar devices, the spectra of Ga-polar LEDs exhibit one main peak with a small shoulder on the low energy side for all the investigated current densities. Therefore, our simulations reproduce well the spectra commonly measured for conventional LEDs. The main peak in the spectra of Ga-polar LEDs originates from the sum of the emissions from QW1, QW2, and QW3, which have all similar emission energies; whereas the slightly red-shifted shoulder is emitted by QW4. 

The main difference between the Ga- and N-polar heterostructures is essentially an inverted stacking sequence of Ga and N atoms in the crystal lattice. Consequently, the total polarization inside the Ga-polar QWs generates a polarization field opposite to the built-in electric field. Therefore, the total electric field in the QWs is weaker than in the N-polar counterpart, and with it also the QCSE. In turn, the weaker QCSE mitigates the red-shift of the emission energies of QW1 and QW4, thus suppressing the formation of the low energy peak observed in the EL spectra of Fig.~\ref{fig:spectra}(a).     

\section{Comparison of experiments with simulations} \label{sec:Comp_exp_sim}

\begin{figure}[t]
\includegraphics*[width=8cm]{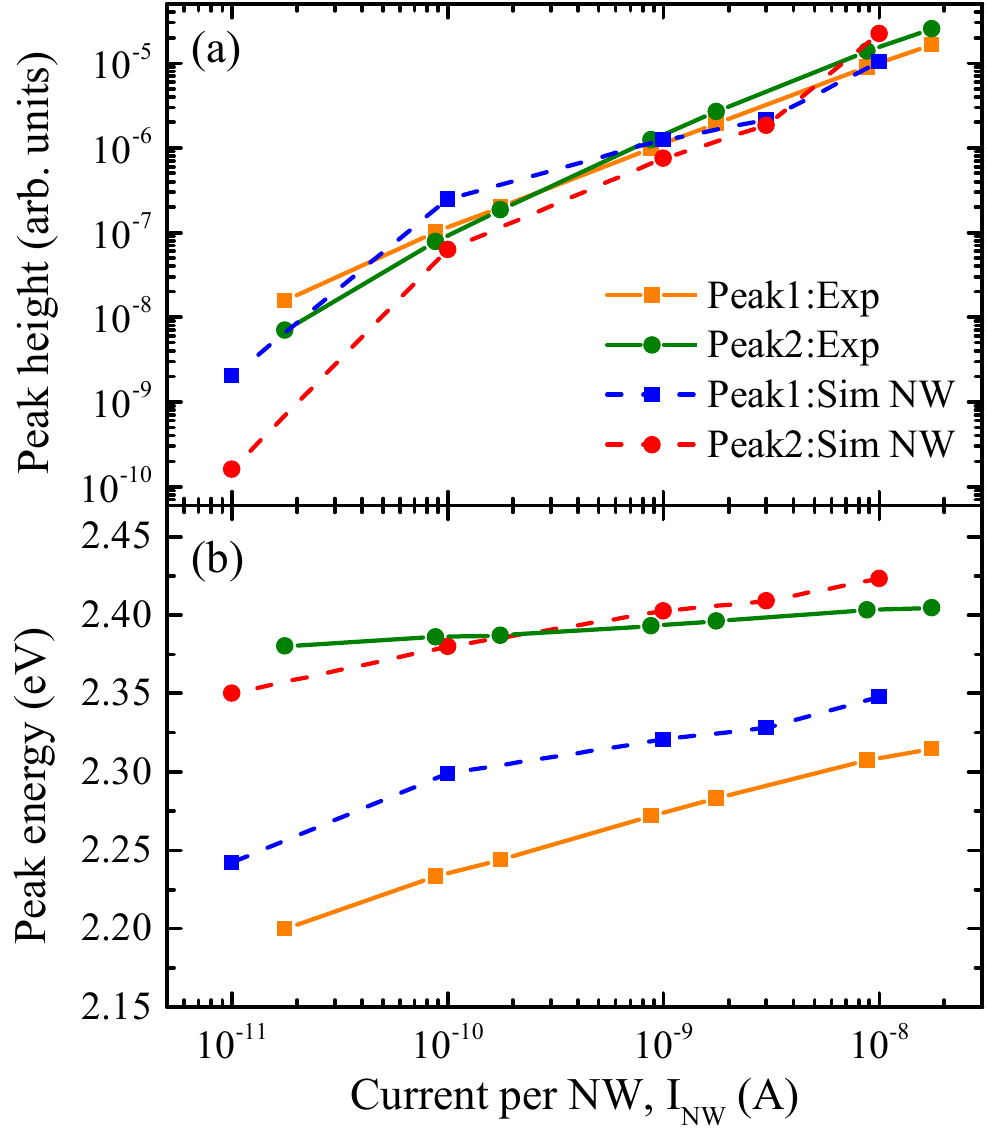} 
\caption{Comparison between simulated and measured optoelectronic properties of the NW-LEDs; see dashed and solid lines, respectively. (a) EL intensity and (b) energy position of the two main peaks plotted \textit{vs} the \emph{average} current per NW.}
\label{fig:comp}
\end{figure}

In this section, we compare in more detail the main emission properties obtained from the simulations of the NW-LEDs with the ones extracted from the experimental measurements. Figures~\ref{fig:comp}(a) and (b) summarize the intensity and energy position of the two main transitions observed in the EL spectra for a range of injected currents that varies over more than three orders of magnitude. These data were obtained by fitting the EL spectra with two Gaussians. For the experimental data, the \emph{average} current per NW ($I_{NW}$) was estimated by using the procedure described in one of our previous publications~\cite{Musolino2014}, namely: $I_{NW}=I/A\,D_{on}$, where $I$, $A$, and $D_{on}$ denote the total current injected into the NW ensemble, the total device area, and the number density of NWs emitting light extracted from EL maps, respectively. Because of the rather big error in the estimation of $D_{on}$, the evaluation of the average current per NW is also affected by a significant uncertainty; further details can be found elsewhere~\cite{Musolino2014,MusolinoPhD,Musolino_NT_2016}.

As shown in Fig.~\ref{fig:comp}(a), the low-energy peak (\emph{i.\,e.}, Peak1) is brighter at low injected currents, but its intensity increases more slowly than the one of Peak2, and eventually the high-energy peak dominates over the low-energy one. The simulations reproduce the peculiar experimental trend found for the EL intensity of NW-LEDs. The agreement between simulated and measured intensities is even quantitatively rather good over almost the entire current range. The only exception are the data points at the lowest currents very close to turn-on, where the behaviour of the LED changes drastically for small variations in current. Therefore, this comparison demonstrates that our simulations can describe very well how the EL intensity emitted by NW-LEDs varies with current.
 
Figure~\ref{fig:comp}(b) presents the evolution with current of the energy positions obtained from the fits of the two main transitions. In the simulations, we varied the In content in order to obtain the best fit to the experimental data related to Peak2, which is the line that would dominate the spectra at typical operating currents. A rather good agreement between simulations and experiments is obtained for an In content in the QWs of 32$\%$. Figures~\ref{fig:comp} (a) and (b) show only the simulated data related to this In concentration. For both, Peak1 and Peak2, the energy shift per current decade, namely the slope of the curves, obtained from the simulations agrees well with the measured value. Furthermore, we note that the energy positions of Peak1 obtained from the simulations are systematically blue-shifted with respect to the experimental data points. An energy difference between simulations and experiments of about 50\,meV is observed. 

The calculated EL spectra exhibit narrower lines than those observed in the experimental spectra. However, there are differences between Peak1 and Peak2. While for Peak1 the average full widths at half maximum (FWHM) obtained from the simulations are roughly nine times smaller than those obtained from the experiments, these are only two times lower for Peak2. The larger experimental values for the widths of the transitions can be explained by structural fluctuations and deviations from the idealized schematic shown in Fig.~\ref{fig:sketch}. The fact that the difference between experimental and simulated FWHM is much larger for Peak1 than for Peak2 implies that these factors are more relevant for the former transition. Thus, it is plausible that also the energy position of Peak1 is reproduced less accurately by the simulations. Note that the broad Peak1 could also result from the superposition of different mechanisms emitting EL. It could partially be caused by the QCSE, as indicated by our simulations, and partially by other effects related to defects, as suggested in case of Ga-polar planer LEDs by varius groups~\cite{Mao_2011,Meneghini_2012}.

As pointed out at the end of section~\ref{sec:methods}, the actual structure of the NWs may be more complex than that sketched in Fig.~\ref{fig:sketch} and considered in our simulations. Furthermore, physical phenomena that are known to be important for the luminescence of such NWs, in particular In segregation and the formation of crystal defects, are not included in the theoretical description. In view of these limitations, the agreement between our experiments and simulations as shown in Fig.~\ref{fig:comp} is rather good. Therefore, we conclude that TiberCAD is suited to describe the main aspects in the physics of NW-LEDs and that our explanation of the behaviour of the EL spectra is reasonable.

\section{Summary}

Comparing measured spectra with simulated data, we identified two overlapping main transitions contributing to the EL spectra of NW-LEDs. Their relative intensities evolve in a peculiar way with injected current. Spatially and spectrally resolved EL maps confirm that the two lines are emitted by the majority of the single nano-LEDs in the ensemble. Such a double-peak structure of the EL is usually not observed in conventional LEDs based on Ga-polar planar (In,Ga)N/GaN heterostructures.

In order to understand the physical origin of the two transitions, we computed strain, electric field, and charge density distributions inside the active region of both planar LEDs and NW-LEDs. On this basis, also the EL spectra of the corresponding devices were simulated. Interestingly, the simulated EL spectra of N-polar heterostructures exhibit a double-peak structure for both the NW and planar case, this feature is not observed for Ga-polar planar LEDs. Therefore, we conclude that the formation of the double-peak structure in the EL spectra is a characteristic of the N-polar LEDs, and originates from a stronger QCSE due to the higher electric field strengths in the QWs close to the n-type base and p-type EBL in these LEDs.

Furthermore, the current dependence of the measured EL intensities of the two transitions is reproduced only in 3D simulations that take into account the NW geometry. It can be explained by the reduction of electric field and QCSE inside the QW in contact with the EBL, due to the strain relaxation occurring at the free sidewalls of the NWs. Moreover, by analysing the 3D charge density distribution inside each QW, we found that an in-plane carrier localization occurs inside the QWs of the NW-LED. The magnitude of this phenomenon differs from QW to QW and we attribute this difference to a not uniform 3D strain distribution inside the active region of the NW-LED, which induces a complex redistribution of electric fields and charge carriers. 

In conclusion, we identified consequences of the strain relaxation induced by the 3D morphology of the NWs for the emission properties of (In,Ga)N/GaN NW-LEDs. Furthermore, our study revealed that the crystal polarity plays an important role in the operation of such LEDs in comparison to their conventional planar counterpart.

\section{Acknowledgement}

The authors gratefully acknowledge O. Brandt  as well as O. Marquardt for constructive critiques to this work, M. Auf der Maur for useful discussions, and W. Anders for his help with the fabrication of the devices. We are also grateful to L. Schrottke for a critical reading of the manuscript. This work was partially supported by the European Commission (project DEEPEN, FP7-NMP-2013-SMALL-7, grant agreement no. 604416).

\bibliography {bibl_5}

\end{document}